\documentstyle[12pt]{article}

\font\twlgot =eufm10 scaled \magstep1
\font\egtgot =eufm8
\font\sevgot =eufm7

\font\twlmsb =msbm10 scaled \magstep1
\font\egtmsb =msbm8
\font\sevmsb =msbm7

\newfam\gotfam

\textfont\gotfam\twlgot
\scriptfont\gotfam\egtgot
\scriptscriptfont\gotfam\sevgot

\newfam\msbfam
\textfont\msbfam\twlmsb
\scriptfont\msbfam\egtmsb
\scriptscriptfont\msbfam\sevmsb
\def\Bbb{\protect\pBbb}
\def\pBbb{\relax\ifmmode\expandafter\Bb\else\typeout{You cann't use
Bbb in text mode}\fi}
\def\Bb #1{{\fam\msbfam\relax#1}}

\def\thebibliography#1{\section*{References}\list
  {[\arabic{enumi}]}{\settowidth\labelwidth{#1}\leftmargin\labelwidth
    \advance\leftmargin\labelsep
    \usecounter{enumi}}
    \def\newblock{\hskip .11em plus .33em minus .07em}
    \sloppy\clubpenalty4000\widowpenalty4000
    \sfcode`\.=1000\relax}

\def\op#1{\mathop{\fam0 #1}\limits}

\newcommand{\id}{{\rm Id\,}}

\newcommand{\Ker}{{\rm Ker\,}}

\newcommand{\beq}{\begin{equation}}
\newcommand{\eeq}{\end{equation}}
\newcommand{\ben}{\begin{eqnarray}}
\newcommand{\een}{\end{eqnarray}}
\newcommand{\be}{\begin{eqnarray*}}
\newcommand{\ee}{\end{eqnarray*}}
\newcommand{\bea}{\begin{eqalph}}
\newcommand{\eea}{\end{eqalph}}

\newcommand{\cT}{{\cal T}}
\newcommand{\cP}{{\cal P}}
\newcommand{\cR}{{\cal R}}
\newcommand{\cL}{{\cal L}}

\newcommand{\cH}{{\cal H}}
\newcommand{\cF}{{\cal F}}

\newcommand{\cS}{{\cal S}}

\newcommand{\cN}{{\cal N}}
\newcommand{\ccG}{{\cal G}}
\newcommand{\bL}{{\bf L}}

\newcommand{\al}{\alpha}

\newcommand{\la}{\lambda}
\newcommand{\La}{\Lambda}
\newcommand{\f}{\phi}
\newcommand{\om}{\omega}
\newcommand{\Om}{\Omega}
\newcommand{\m}{\mu}

\newcommand{\g}{\gamma}
\newcommand{\G}{\Gamma}

\newcommand{\th}{\theta}

\newcommand{\si}{\sigma}
\newcommand{\w}{\wedge}

\newcommand{\wh}{\widehat}
\newcommand{\ol}{\overline}
\newcommand{\dr}{\partial}
\newcommand{\ar}{\op\longrightarrow}

\newcommand{\ot}{\otimes}

\newcounter{theorem}
\newcounter{remark}
\newcounter{proposition}
\newcounter{lemma}
\newcounter{corollary}
\newcounter{definition}

\def\theremark{\arabic{remark}}

\def\thedefinition{\arabic{definition}}

\newenvironment{theo}{\refstepcounter{definition} \medskip
\noindent{\bf Theorem \thedefinition.} }{\medskip}
\newenvironment{prop}{\refstepcounter{definition} \medskip
\noindent{\bf Proposition \thedefinition.} }{\medskip}

\newcommand{\mar}[1]{}

\begin{document}

\hbox{}

\begin{center}

{\large\bf POLYSYMPLECTIC HAMILTONIAN FORMALISM AND SOME QUANTUM
OUTCOMES}

\bigskip

{\sc G.Giachetta, L.Mangiarotti$^1$ and G.Sardanashvily$^2$}

\medskip

$^1$ Department of Mathematics and Informatics, University of
Camerino, 62032 Camerino (MC), Italy

$^2$ Department of Theoretical Physics, Moscow State
University, 117234 Moscow, Russia

\end{center}

\bigskip
\bigskip

{\small

Covariant (polysymplectic) Hamiltonian field theory is
formulated as a particular Lagrangian theory on a
polysymplectic phase space that enables one to quantize it in
the framework of familiar quantum field theory. 

}

\section{Introduction}

The Hamiltonian counterpart of first-order
Lagrangian formalism on a fibre bundle $Y\to X$ has been
rigorously developed since the 1970s in the 
multisymplectic, polysymplectic and Hamilton -- De Donder
variants (see
\cite{ech00,ech04,book,jpa99,hel,krupk2,leon,leon2} and
references therein). If $X=\Bbb R$, we are in the case of
time-dependent mechanics \cite{book98}.

The relations between multisymplectic,
polysymplectic, Hamilton -- De Donder and Lagrangian formalisms
on
$Y\to X$ are briefly the following.

$\bullet$ The multisymplectic phase space is  
the homogeneous Legendre bundle
\mar{N41}\beq
Z_Y= T^*Y\w(\op\w^{n-1}T^*X),  \label{N41}
\eeq
coordinated by $(x^\la,y^i,p^\la_i,p)$. It is
endowed with the canonical exterior form
\beq
\Xi_Y= p\om + p^\la_i dy^i\w\om_\la, \label{N43}
\eeq
whose exterior differential $d\Xi_Y$ is the canonical
multisymplectic form, which belongs to the class of
multisymplectic forms in the sense of Martin \cite{cantr,mart}. 

$\bullet$ The homogeneous Legendre bundle (\ref{N41}) 
is the trivial one-dimensional
bundle
\mar{N41'}\beq
\zeta:Z_Y\to \Pi \label{N41'}
\eeq
over the Legendre bundle
\mar{00}\beq
\Pi=\op\w^nT^*X\op\ot_YV^*Y\op\ot_YTX=V^*Y\w(\op\w^{n-1}T^*X),
\label{00}
\eeq
coordinated by $(x^\la,y^i,p^\m_i)$. Being provided with the
canonical polysymplectic form 
\mar{406}\beq
\Om =dp_i^\la\w dy^i\w \om\ot\dr_\la, \label{406}
\eeq
the Legendre bundle $\Pi$ is the momentum phase space of
polysymplectic Hamiltonian formalism.
A Hamiltonian $\cH$ on $\Pi$ is defined as a section $p=-\cH$ of
the bundle $\zeta$ (\ref{N41'}). The pull-back of $\Xi_Y$ onto
$\Pi$ by a Hamiltonian $\cH$ is a Hamiltonian form
\mar{b418}\beq
 H=\cH^*\Xi_Y= p^\la_i dy^i\w \om_\la -\cH\om.  \label{b418}
\eeq

In the case of mechanics,  $Z_Y=TY$ and $\Pi=VY$ are the
homogeneous momentum phase space and the momentum phase space of
time-dependent mechanics on $Y\to \Bbb R$, respectively.
Accordingly, $H$ (\ref{b418}) is the well-known
integral invariant of Poincar\'e--Cartan.

From the mathematical viewpoint, an
essential advantage of a multisymplectic formalism is that the
multisymplectic form is an exterior form. In physical
applications, one however meets an additional
variable $p$ which is the energy one in homogeneous
time-dependent mechanics.

It should be emphasized that multisymplectic and polysymplectic
formalisms need not be related to Lagrangian one. In
contrast with them, Hamilton -- De Donder formalism necessarily
describes Lagrangian systems as follows.

Let us consider a first order Lagrangian
\mar{cmp1}\beq
L=\cL\om: J^1Y\to\op\w^nT^*X, \quad \om=dx^1\w\cdots\w dx^n, \quad
n=\dim X, \label{cmp1}
\eeq
on $J^1Y$, the Euler--Lagrange equations
\mar{b327} \beq
(\dr_i- d_\la\dr^\la_i)\cL=0, \label{b327}
\eeq
and the Poincar\'e--Cartan form
\mar{303}\beq
 H_L=L +\pi^\la_i\th^i\w\om_\la, \quad \pi^\la_i=\dr^\la_i\cL,
 \quad \om_\la=\dr_\la\rfloor\om.
\label{303}
\eeq
The latter is both the particular Lepagean equivalent of a
Lagrangian $L$ (\ref{cmp1}) and that of the Lagrangian
\mar{cmp80}\beq
 \ol L=\wh h_0(H_L) = (\cL + (\wh y_\la^i -
y_\la^i)\pi_i^\la)\om, \qquad \wh h_0(dy^i)=\wh y^i_\la dx^\la,
\label{cmp80}
\eeq
on the repeated jet manifold $J^1J^1Y$. Its Euler--Lagrange
equations are the Cartan equations
\mar{b336}\beq
\dr_i^\la\pi_j^\m(\wh y_\m^j - y_\m^j)=0, \quad
 \dr_i \cL - \wh
d_\la\pi_i^\la + (\wh y_\la^j - y_\la^j)\dr_i\pi_j^\la=0.
\label{b336}
\eeq

$\bullet$ The Poincar\'e--Cartan form (\ref{303})
yields the Legendre morphism
\be
\wh H_L: J^1Y\op\to_Y Z_Y, \quad (p^\m_i, p)\circ\wh H_L
=(\pi^\m_i, \cL-\pi^\m_i y^i_\m ),
\ee
of $J^1Y$ to the homogeneous Legendre bundle $Z_Y$ (\ref{N41}).
Let its image $Z_L=\wh H_L(J^1Y)$ be an imbedded subbundle
$i_L:Z_L\hookrightarrow Z_Y$ of $Z_Y\to Y$. Then it is provided
with the pull-back De Donder form $\Xi_L=i^*_L\Xi_Y$. 
The Hamilton -- De Donder
equations for sections $\ol r$ of $Z_L\to X$ are written as
\mar{N46}\beq
\ol r^*(u\rfloor d\Xi_L)=0, \label{N46}
\eeq
where $u$ is an arbitrary vertical vector field on $Z_L\to X$.
 Let the Legendre morphism
$\wh H_L$ be a submersion. Then one can show that a section $\ol
s$ of
$J^1Y\to X$ is a solution of the Cartan equations (\ref{b336})
iff $\wh H_L\circ\ol s$ is a solution of the Hamilton--De
Donder equations (\ref{N46}). In a general setting, one can
consider different Lepagean forms in order to develop Hamilton
-- De Donder formalism \cite{krupk1,krupk2}. 

$\bullet$ The relation between polysymplectic Hamiltonian and 
Lagrangian formalisms is based on the fact that  any Lagrangian
$L$ yields the Legendre map
\mar{m3}\beq
\wh L: J^1Y\ar_Y \Pi, \quad p^\la_i\circ\wh L=\dr^\la_i\cL,
\label{m3}
\eeq
whose image $N_L=\wh L(J^1Y)$ is called the Lagrangian constraint
space. Conversely, any Hamiltonian $\cH$ defines the Hamiltonian
map
\mar{415}\beq
\wh H: \Pi\ar_Y J^1Y, \quad y_\la^i\circ\wh H=\dr^i_\la\cH.
\label{415}
\eeq
A Hamiltonian $\cH$ on $\Pi$ is said to be associated to a
Lagrangian $L$ on $J^1Y$ if it satisfies the relations
\mar{2.30a,b}\ben
&& p^\m_i=\dr^\m_i\cL
(x^\m,y^i,\dr^j_\la\cH),
\label{2.30a} \\
&&p^\m_i\dr^i_\m\cH-\cH=\cL(x^\m,y^j,\dr^j_\la\cH).
\label{2.30b}
\een
If an associated Hamiltonian $\cH$ exists, the Lagrangian
constraint space $N_L$ is given by the coordinate relations
(\ref{2.30a}) and $\wh L\circ \wh H$ is a projector 
of $\Pi$ onto
$N_L$.

Lagrangian and polysymplectic Hamiltonian formalisms are
equivalent in the case of hyperregular Lagrangians. 
The
key point is that a degenerate Lagrangian admits different
associated Hamiltonians, if any. At the same time, 
there is a comprehensive relation between
these formalisms in the case of almost-regular Lagrangians. 
Recall that a Lagrangian $L$ is called almost-regular if the
Lagrangian constraint space  is a closed imbedded subbundle
$i_N:N_L\to \Pi$ of the Legendre bundle $\Pi\to Y$ and the
surjection $\wh L:J^1Y\to N_L$ is a fibred manifold possessing
connected fibres. In particular, the Poincar\'e--Cartan form
(\ref{303}) is the pull-back $H_L=\wh L^*H$ of the Hamiltonian
form $H$ (\ref{b418}) for any associated Hamiltonian $\cH$.

Now let us focus on polysymplectic Hamiltonian formalism
\cite{book,jpa99}. Bearing in mind its quantization, we
formulate it as particular Lagrangian formalism on the
Legendre bundle $\Pi$ (\ref{00}).

\section{Polysymplectic Hamiltonian dynamics}

For every Hamiltonian form $H$ (\ref{b418}), there
exists a connection
\mar{cmp33}\beq
\g =dx^\la\otimes(\dr_\la +\g^i_\la\dr_i +\g^\m_{\la i}\dr^i_\m)
\label{cmp33}
\eeq
on $\Pi\to X$ such that
\mar{cmp3}\beq
\g\rfloor\Om= dH, \qquad \g^i_\la=\dr^i_\la\cH, \qquad \g^\la_{\la
i}= -\dr_i\cH. \label{cmp3}
\eeq
The connection (\ref{cmp33}), called the Hamiltonian connection,
yields the first order dynamic Hamilton equations on $\Pi$ given
by the closed submanifold
\mar{b4100}\beq
y^i_\la=\dr^i_\la\cH, \qquad  p^\la_{\la i}=-\dr_i\cH
\label{b4100}
\eeq
of the jet manifold $J^1\Pi$ of $\Pi\to X$.

A polysymplectic Hamiltonian system on $\Pi$ is equivalent to
the above mentioned particular Lagrangian system on $\Pi$ as
follows.

\begin{prop} \label{m11} \mar{m11}
The Hamilton equations (\ref{b4100}) are equivalent to the
Euler--Lagrange equations for the first-order Lagrangian
\mar{m5}\beq
L_\cH=h_0(H)=\cL_\cH\om= (p^\la_i y^i_\la-\cH)\om. \label{m5}
\eeq
\end{prop}

Let $i_N:N\to \Pi$ be a closed imbedded subbundle of the Legendre
bundle $\Pi\to Y$ which is regarded as a constraint space of a
polysymplectic Hamiltonian field system with a Hamiltonian $\cH$.
Let $H_N=i^*_NH$ be the pull-back of the Hamiltonian form $H$
(\ref{b418}) onto $N$. This form defines the constrained
Lagrangian
\mar{cmp81}\beq
L_N=h_0(H_N)=(J^1i_N)^*L_\cH \label{cmp81}
\eeq
on the jet manifold $J^1N_L$ of the fibre bundle $N_L\to X$. The
Euler--Lagrange equations for this Lagrangian are called the
constrained Hamilton equations.

In fact, the Lagrangian $L_\cH$ (\ref{m5}) is the pull-back onto
$J^1\Pi$ of the horizontal form $L_\cH$ on the bundle product
$\Pi\op\times_Y J^1Y$ by the canonical map 
\be
J^1\Pi\to
\Pi\op\times_Y J^1Y.
\ee
Therefore, the constrained Lagrangian $L_N$
(\ref{cmp81}) is simply the restriction of $L_\cH$ to
$N\op\times_Y J^1Y$.

\begin{prop} \label{m10} \mar{m10}
A section $r$ of $\Pi\to X$ is a solution of the Hamilton
equations (\ref{b4100}) iff it satisfies the condition
\be
r^*(u_\Pi\rfloor dH)= 0
\ee
for any vertical vector field
$u_\Pi$ on
$\Pi\to X$.
\end{prop}

\begin{prop} \label{m12} \mar{m12}
A section $r$ of the fibre bundle $N\to X$ is a solution of
constrained Hamilton equations iff it satisfies the condition
$r^*(u_N\rfloor dH)= 0$ for any vertical vector field $u_N$ on
$N\to X$.
\end{prop}

Propositions \ref{m10} and \ref{m12} result in the following.

\begin{prop} \label{y23} \mar{y23}
Any solution of the Hamilton equations (\ref{b4100}) which lives
in the constraint manifold $N$ is also a solution of the
constrained Hamilton equations on $N$.
\end{prop}

Forthcoming Theorems \ref{3.23} - \ref{3.01} establish the
above mentioned relation between Lagrangian and polysymplectic
Hamiltonian formalisms in the case of almost-regular
Lagrangians.

\begin{theo}\label{3.23} \mar{3.23}
Let $L$ be an almost-regular Lagrangian and $\cH$ an associated
Hamiltonian. Let a section $r$ of $\Pi\to X$ be a  solution of the
Hamilton equations (\ref{b4100}) for $\cH$. If $r$ lives in the
Lagrangian constraint manifold $N_L$, then $s=\pi_Y\circ r$
satisfies the Euler--Lagrange equations (\ref{b327}) for $L$,
while $\ol s=\wh H\circ r$ obeys the Cartan equations
(\ref{b336}). Conversely, let $\ol s$ be a solution of the Cartan
equations (\ref{b336}) for $L$. If $\cH$ satisfies the relation
\be
\wh H\circ \wh L\circ \ol s=J^1(\pi^1_0\circ\ol s),
\ee
the section $r=\wh L\circ \ol s$ of the Legendre bundle $\Pi\to X$
is a solution of the Hamilton equations (\ref{b4100}) for $\cH$.
\end{theo}

If an almost-regular Lagrangian admits associated Hamiltonians
$\cH$, they define a unique constrained Lagrangian $L_N=h_0(H_N)$
(\ref{cmp81}) on the jet manifold $J^1N_L$ of the fibre bundle
$N_L\to X$. Basing on Proposition \ref{y23} and Theorem
\ref{3.23}, one can prove the following.

\begin{theo}\label{3.01} \mar{3.01} Let an almost-regular Lagrangian
$L$ admit associated Hamiltonians. A section
$\ol s$ of the jet bundle $J^1Y\to X$ is a solution of the Cartan
equations for $L$ iff
$\wh L\circ \ol s$ is a solution of  the constrained Hamilton equations.
In particular, any solution $r$ of the constrained Hamilton equations
provides the solution $\ol s=\wh H\circ r$ of the Cartan equations.
\end{theo}

Thus, one can associate to an almost-regular Lagrangian
(\ref{cmp1}) a unique constrained Lagrangian system on the
constraint Lagrangian manifold $N_L$ (\ref{2.30a}).

\section{Quadratic degenerate systems}

Quadratic Lagrangians provide a most physically relevant example
of degenerate Lagrangian systems.

Let us consider a  quadratic Lagrangian
\mar{N12}\beq
\cL=\frac12
a^{\la\m}_{ij} y^i_\la y^j_\m + b^\la_i y^i_\la + c, \label{N12}
\eeq
where $a$, $b$ and $c$ are local functions on $Y$. The associated
Legendre map (\ref{m3}) reads
\mar{N13}\beq
p^\la_i\circ\wh L= a^{\la\m}_{ij} y^j_\m +b^\la_i. \label{N13}
\eeq

Let a Lagrangian $L$ (\ref{N12}) be almost-regular, i.e.,
the matrix function $a$ is a linear bundle morphism
\mar{m38}\beq
a: T^*X\op\ot_Y VY\to \Pi, \quad p^\la_i=a^{\la\m}_{ij} \ol
y^j_\m, \label{m38}
\eeq
of constant rank, where $(x^\la,y^i,\ol y^i_\la)$ are coordinates
on $T^*X\op\ot_Y VY$. Then the Lagrangian constraint space $N_L$
(\ref{N13}) is an affine subbundle of $\Pi\to Y$. Hence, $N_L\to
Y$ has a global section. Let us assume that it is the canonical
zero section $\wh 0(Y)$ of $\Pi\to Y$. The kernel of the Legendre
map (\ref{N13}) is also an affine subbundle of the affine jet
bundle $J^1Y\to Y$. Therefore, it admits a global section
\mar{N16}\beq
\G: Y\to \Ker\wh L\subset J^1Y, \qquad
a^{\la\m}_{ij}\G^j_\m + b^\la_i =0,  \label{N16}
\eeq
which is a connection on $Y\to X$. With $\G$, the Lagrangian
(\ref{N12}) is brought into the form
\mar{y47}\beq
\cL=\frac12 a^{\la\m}_{ij} (y^i_\la -\G^i_\la)(y^j_\m-\G^j_\m) +c'.
\label{y47}
\eeq

\begin{theo}\label{04.2}  There exists a linear bundle
morphism
\mar{N17}\ben
&& \si: \Pi\op\to_Y T^*X\op\otimes_YVY, \quad \ol y^i_\la\circ\si
=\si^{ij}_{\la\m}p^\m_j, \label{N17}\\
&& a\circ\si\circ a=a, \qquad
a^{\la\mu}_{ij}\si^{jk}_{\mu\al}a^{\al\nu}_{kb}=a^{\la\nu}_{ib}.
\label{N45}
\een
\end{theo}

The morphism $\si$ (\ref{N17}) is not unique, but it falls into
the sum $\si=\si_0+\si_1$ such that
\mar{N21}\beq
\si_0\circ a\circ \si_0=\si_0, \qquad a\circ\si_1=\si_1\circ a=0, \label{N21}
\eeq
where $\si_0$ is uniquely defined. The equalities (\ref{N16}) and
(\ref{N45}) give the relation 
\be
(a\circ\si_0)^{\la j}_{i\m}
b^\m_j=b^\la_i.
\ee

\begin{theo} \label{m15} \mar{m15}
There are the splittings
\mar{N18,20}\ben
&& J^1Y=\Ker\wh L\op\oplus_Y{\rm Im}(\si_0\circ
\wh L), \label{N18} \\
&& y^i_\la=\cS^i_\la+\cF^i_\la= [y^i_\la
-\si_0{}^{ik}_{\la\al} (a^{\al\m}_{kj}y^j_\m + b^\al_k)]+
[\si_0{}^{ik}_{\la\al} (a^{\al\m}_{kj}y^j_\m + b^\al_k)], \nonumber\\
&& \Pi=\Ker\si_0 \op\oplus_Y N_L, \label{N20} \\
&& p^\la_i = \cR^\la_i+\cP^\la_i= [p^\la_i -
a^{\la\m}_{ij}\si_0{}^{jk}_{\m\al}p^\al_k] +
[a^{\la\m}_{ij}\si_0{}^{jk}_{\m\al}p^\al_k]. \nonumber
\een
\end{theo}

The relations (\ref{N21}) lead to the equalities
\mar{m25}\beq
\si_0{}^{jk}_{\m\al}\cR^\al_k=0, \quad
\si_1{}^{jk}_{\m\al}\cP^\al_k=0, \quad \cR^\la_i\cF^i_\la=0.
\label{m25}
\eeq

By virtue of the equalities (\ref{N21}) and the relation
\mar{y75}\beq
\cF^i_\m=(\si_0\circ a)^{i\la}_{\m
j}(y^j_\la-\G^j_\la), \label{y75}
\eeq
the Lagrangian (\ref{N12}) takes the form
\mar{cmp31}\beq
L=\cL\om, \qquad \cL=\frac12 a^{\la\m}_{ij}\cF^i_\la\cF^j_\m +c'. \label{cmp31}
\eeq
It admits a set of associated Hamiltonians
\mar{N22}\beq
\cH_\G=(\cR^\la_i+\cP^\la_i)\G^i_\la
+\frac12
\si_0{}^{ij}_{\la\m}\cP^\la_i\cP^\m_j
+\frac12\si_1{}^{ij}_{\la\m}\cR^\la_i\cR^\m_j -c'\label{N22}
\eeq
indexed by connections $\G$ (\ref{N16}). Accordingly, the Lagrangian constraint
manifold (\ref{N13}) is characterized by the equalities
\mar{bv1}\beq
\cR^\la_i=p^\la_i -
a^{\la\m}_{ij}\si_0{}^{jk}_{\m\al}p^\al_k=0. \label{bv1}
\eeq

Given a Hamiltonian $\cH_\G$, the corresponding Lagrangian
(\ref{m5}) on $\Pi\op\times_Y J^1Y$ reads
\mar{m16}\beq
\cL_{\cH_\G}=\cR^\la_i(\cS^i_\la-\G^i_\la) +\cP^\la_i\cF_\la^i
-\frac12\si_0{}^{ij}_{\la\m}\cP^\la_i\cP^\m_j -
\frac12\si_1{}^{ij}_{\la\m} \cR^\la_i \cR^\m_j+ c'. \label{m16}
\eeq
Its restriction (\ref{cmp81}) to the constraint manifold $N_L\op\times_YJ^1Y$
is
\mar{bv2}\beq
L_N=\cL_N\om, \qquad \cL_N=\cP^\la_i\cF_\la^i
-\frac12\si_0{}^{ij}_{\la\m}\cP^\la_i\cP^\m_j + c'. \label{bv2}
\eeq
It is independent of the choice of a Hamiltonian (\ref{N22}).

The Hamiltonian $\cH_\G$ yields the Hamiltonian map $\wh H_\G$ and
the projector
\be
\cT=\wh L\circ \wh H_\G,\qquad p^\la_i\circ \cT=\cT^{\la j}_{i
\m}p^\m_j=a^{\la\nu}_{ik}\si_0{}^{kj}_{\nu\m}
p^\m_j=\cP^\la_i,\label{m30}
\ee
of $\Pi$ onto its summand $N_L$ in the decomposition (\ref{N20}).
It is a linear morphism over $\id Y$. Therefore, $\cT:\Pi\to N_L$
is a vector bundle. Let us consider the pull-back
\mar{m32}\beq
L_\Pi=\cT^*L_N=\cL_\Pi\om, \qquad \cL_\Pi=\cP^\la_i\cF_\la^i
-\frac12\si_0{}^{ij}_{\la\m}\cP^\la_i \cP^\m_j + c', \label{m32}
\eeq
of the constrained
Lagrangian $L_N$ (\ref{bv2}) onto $\Pi\op\times_Y J^1Y$.

\section{Quantization}

In order to quantize covariant Hamiltonian systems, one usually
attempts to construct the multisymplectic generalization of a
Poisson bracket \cite{lop,for,kanat,kanat2}. In a different
way, we suggested to quantize covariant (polysymplectic)
Hamiltonian field theory in path integral terms \cite{sard94}.
This quantization scheme has been modified in order to quantize
a polysymplectic Hamiltonian
system with a Hamiltonian $\cH$ on $\Pi$ as a Lagrangian system
with the Lagrangian $L_\cH$ (\ref{m5}) in the framework of
familiar quantum field theory \cite{bashk,bashk2}.

If there is no constraint and the matrix
\be
\dr^2\cH/\dr p^\m_i\dr p^\nu_j=-\dr^2\cL/\dr p^\m_i\dr p^\nu_j
\ee
is positive-definite and non-degenerate on an Euclidean
space-time, this quantization is given by the generating
functional
\mar{m2}\beq
Z=\cN^{-1}\int\exp\{\int(\cL_\cH +\Lambda + iJ_iy^i+iJ^i_\m
p^\m_i) \om \}\op\prod_x [dp(x)][dy(x)] \label{m2}
\eeq
of Euclidean Green functions, where $\Lambda$ comes from the
normalization condition
\be
\int \exp\{\int(\frac12\dr_\m^i\dr_\nu^j\cL_\cH p^\m_i
p^\nu_j+\La)dx\}\op\prod_x[dp(x)]=1.
\ee

A constrained Hamiltonian system on a constraint manifold $N$ can
be quantized as a Lagrangian system with the pull-back Lagrangian
$L_N$. Furthermore, a closed imbedded constraint submanifold $N$
of $\Pi$ admits an open neighbourhood $U$ which is a fibred
manifold $U\to N$. If $\Pi$ is a fibred manifold $\pi_N:\Pi\to N$
over $N$, it is often convenient to quantize a Lagrangian system
on $\Pi$ with the pull-back Lagrangian $L_\Pi=\pi_N^*L_N$. Since
this Lagrangian possesses gauge symmetries, BV
(Batalin--Vilkoviski) quantization can be called into play
\cite{bat,gom}.

For instance, BV quantization can be applied to Hamiltonian
systems associated to Lagrangian field systems with quadratic
Lagrangians $L$ (\ref{N12}). If this Lagrangian is hyperregular
(i.e., the matrix function $a$ is non-degenerate), there exists a
unique associated Hamiltonian system whose Hamiltonian $\cH$ is
quadratic in momenta $p^\m_i$, and so is the Lagrangian $\cL_\cH$
(\ref{m5}). If the matrix function $a$ is positive-definite on an
Euclidean space-time, the generating functional (\ref{m2}) is a
Gaussian integral of momenta $p^\m_i(x)$. Integrating $Z$ with
respect to $p^\m_i(x)$, one restarts the generating functional of
quantum field theory with the original Lagrangian $L$ (\ref{N12}).
Using the BV quantization procedure, this result is generalized to
field theories with almost-regular Lagrangians $L$ (\ref{N12}),
e.g., Yang--Mills gauge theory.

The Lagrangian $L_\Pi$ (\ref{m32}) possesses gauge symmetries. By
gauge transformations are meant automorphisms $\Phi$ of the
composite fibre bundle $\Pi\to Y\to X$ over bundle automorphisms
$\f$ of $Y\to X$ over $\id X$. Such an automorphism $\Phi$ gives
rise to the automorphism $(\Phi, J^1\f)$ of the composite fibre
bundle
\be
\Pi\op\times_Y J^1Y\to Y\to X.
\ee
An automorphism $\Phi$ is said to be a gauge symmetry of the
Lagrangian 
$L_\Pi$ if 
\be
(\Phi,J^1\f)^*L_\Pi=L_\Pi.
\ee
If the
Lagrangian (\ref{N12}) is degenerate, the group $G$ of gauge
symmetries of the Lagrangian $L_\Pi$ (\ref{m32}) is never
trivial. Indeed, any vertical automorphism of the vector bundle
$\Ker \si_0\to Y$ in the decomposition (\ref{N20}) is obviously
a gauge symmetry of the Lagrangian $L_\Pi$ (\ref{m32}). The
gauge group
$G$ acts on the space $\Pi(X)$ of sections of the Legendre bundle
$\Pi\to X$. For the purpose of quantization, it suffices to consider a
subgroup $\ccG$ of $G$  which acts freely on
$\Pi(X)$ and satisfies the relation 
\be
\Pi(X)/\ccG=\Pi(X)/G.
\ee
Moreover, we need one-parameter subgroups of $\ccG$. Their
infinitesimal generators are represented by
projectable vector fields
\mar{y20}\beq
u_\Pi=u^i(x^\m,y^j)\dr_i + u^\la_i(x^\m,y^j,p^\m_j)\dr^i_\la \label{y20}
\eeq
on the Legendre bundle $\Pi\to Y$ which give rise to the vector fields
\mar{m47}\beq
\ol u=u^i\dr_i + u_i^\la\dr^i_\la +d_\la u^i\dr_i^\la, \qquad
d_\la=\dr_\la +y_\la^i\dr_i, \label{m47}
\eeq
on $\Pi\op\times_Y J^1Y$. A Lagrangian $L_\Pi$ is invariant under a
one-parameter group of gauge transformations iff its Lie derivative
\be
\bL_{\ol u}L_\Pi=\ol u(\cL_\Pi)\om
\ee
along the infinitesimal generator $\ol u$ (\ref{m47}) of this group vanishes.

Any vertical vector field $u$ on $Y\to X$ gives rise to the vector
field
\mar{y41}\beq
u_\Pi=u^i\dr_i - \dr_ju^i p^\la_i\dr^j_\la  \label{y41}
\eeq
on the Legendre bundle $\Pi$ and to the vector field
\mar{y40}\beq
\ol u_\Pi=u^i\dr_i - \dr_ju^i p^\la_i\dr^j_\la +d_\la u^i\dr^\la_i \label{y40}
\eeq
on $\Pi\op\times_Y J^1Y$.

Let us assume that the one-parameter gauge group with the
infinitesimal generators $u$ preserves the splitting (\ref{N18}),
i.e., $u$  obey the condition
\mar{y49'}\beq
u^k\dr_k(\si_0{}^{im}_{\la\nu}a^{\nu\m}_{mj})+
\si_0{}^{im}_{\la\nu}a^{\nu\m}_{mk}\dr_ju^k -
\dr_ku^i\si_0{}^{km}_{\la\nu}a^{\nu\m}_{mj} =0. \label{y49'}
\eeq

\begin{prop} \label{y41'} \mar{y41'}
If the condition (\ref{y49'}) holds, the vector field $u_\Pi$ (\ref{y41}) is an
infinitesimal gauge symmetry of the
Lagrangian $L_\Pi$ (\ref{m32}) iff $u$ is
an infinitesimal gauge symmetry of the Lagrangian $L$ (\ref{cmp31}).
\end{prop}

\end{document}